\documentclass[aps,prc,reprint,amsmath,amssymb,superscriptaddress,nofootinbib,longbibliography]{revtex4-2}
\usepackage{graphicx}
\usepackage{bm}
\usepackage{xcolor}
\usepackage{dcolumn}
\usepackage{tikz}
\usepackage{soul}
\usetikzlibrary{arrows.meta, positioning, calc}
\definecolor{accent}{HTML}{3B9AB2}

\definecolor{newred}{HTML}{C27D38}

\newcommand{\pjm}{p^{J}_{\mathrm{min}}}
\newcommand{\Teff}{T_{\mathrm{eff}}}

\usepackage[colorlinks=true, linkcolor=blue, citecolor=blue, urlcolor=blue]{hyperref}

\begin{document}

\title{Electromagnetic probes of concurrent minijet-hydrodynamics evolution}

\author{Soham Banerjee}
\email{soham.banerjee@niser.ac.in}
\affiliation{School of Physical Sciences, National Institute of Science Education and Research, HBNI, Jatni 752050, India}
\affiliation{Homi Bhabha National Institute, Training School Complex, Anushaktinagar, Mumbai 400094, India}

\author{Mayank Singh}
\email{mayank.singh@vanderbilt.edu}
\affiliation{Department of Physics and Astronomy, Vanderbilt University, Nashville, TN 37240, USA}

\author{Charles Gale}
\email{charles.gale@mcgill.ca}
\affiliation{Department of Physics, McGill University, 3600 University Street, Montreal, QC H3A 2T8, Canada}

\author{Sangyong Jeon}
\email{sangyong.jeon@mcgill.ca}
\affiliation{Department of Physics, McGill University, 3600 University Street, Montreal, QC H3A 2T8, Canada}

\author{Daniel Pablos}
\email{pablosdaniel@uniovi.es}
\affiliation{Departamento de F\'{\i}sica, Universidad de Oviedo,
Avda. Federico Garc\'{\i}a Lorca 18, 33007 Oviedo, Spain}
\affiliation{Instituto Universitario de Ciencias y Tecnolog\'{\i}as Espaciales de Asturias (ICTEA), Calle de la Independencia 13, 33004 Oviedo, Spain}

\author{Jean-François Paquet}
\email{jean-francois.paquet@vanderbilt.edu}
\affiliation{Department of Physics and Astronomy, Vanderbilt University, Nashville, TN 37240, USA}


\begin{abstract}

We investigate direct photon and dilepton production in a concurrent minijet-hydrodynamics framework that consistently incorporates event-by-event initial-state fluctuations and dynamical minijet-medium interactions. Produced in initial hard scatterings, minijets thermalize more slowly than the bulk medium and persist into the hydrodynamic stage, where they deposit energy and momentum into the evolving plasma, generating wakes that modify its spacetime evolution. 
Minijets modify the hadronic multiplicity and momentum anisotropy in a manner that can largely be absorbed by adjusting the shear viscosity and the normalization of the system's initial energy-momentum tensor.
In contrast, electromagnetic probes are emitted throughout the evolution and remain directly sensitive to the underlying spacetime history. 
We perform a comprehensive calculation of direct photon and dilepton production with the minijets and bulk medium evolved simultaneously, using PYTHIA to produce minijets, and describing the underlying plasma with IP-Glasma initial conditions, hydrodynamic evolution with MUSIC and hadronic transport with UrQMD. We find that the minijet-induced modification of the hydrodynamic evolution produces measurable effects in both photon and dilepton observables, establishing electromagnetic radiation as a sensitive probe of the interplay between hard partons and the quark-gluon plasma.
\end{abstract}

\maketitle

\section{Introduction}\label{sec:intro}

Relativistic heavy-ion collisions at the Relativistic Heavy Ion Collider (RHIC) and the Large Hadron Collider (LHC) produce the quark-gluon plasma (QGP), a deconfined state of QCD matter that filled the universe during the first few microseconds after the Big Bang. The properties and dynamics of the QGP are governed by Quantum Chromodynamics (QCD), the fundamental theory of the strong interaction. The spacetime evolution of the matter created in these collisions is successfully described by models based on relativistic viscous hydrodynamics~\cite{Heinz:2013th,Gale:2013da,Shen:2020mgh,DerradideSouza:2015kpt}. Hydrodynamics becomes applicable remarkably soon after the initial nuclear impact, at times of order 1 fm. Understanding how the system evolves so rapidly into a regime where a hydrodynamic description is valid remains an active topic of research~\cite{Berges:2020fwq,Schlichting:2019abc}.

Within the bottom-up thermalization scenario~\cite{Baier:2000sb}, QCD matter is understood to thermalize through the rapid multiplication of gluons via radiation. These gluons then interact among themselves, bringing the system closer to thermalization. The initial gluons are produced at the saturation scale $Q_s$. In addition, hard scatterings during the initial collisions generate back-to-back partons with transverse momenta ($p_T$) well above the saturation scale. These highly energetic parton cascades evolve into jets, which can be identified experimentally. Very hard jets ($p_T \geq 20$ GeV) are relatively rare and usually an event has at most one pair of back-to-back hard jets. Some hard scatterings produce parton pairs at intermediate energy ($Q_s \leq p_T \leq 20$ GeV) which are not immediately thermalized. These particles, commonly referred to as minijets, have been studied extensively in Refs.~\cite{Iancu:2015uja,Zhao:2021vmu,Mehtar-Tani:2022zwf,BarreraCabodevila:2022jhi,Zhou:2024ysb,Boguslavski:2025ylx,Soudi:2026fls}. The more energetic minijets may remain out of equilibrium at the onset of hydrodynamic evolution.
They lose energy as they traverse the bulk hydrodynamic medium and, in turn, generate wakes in the fluid. As their production cross sections are much higher than those for high-$p_T$ jets, there can be multiple minijets in a single collision event.

A concurrent minijet-hydrodynamics framework was developed in~\cite{Pablos:2022piv} to investigate the impact of minijets on the evolution of the bulk medium. In particular, the study analyzed the effect of minijet-induced wakes on the hydrodynamic evolution and soft hadronic observables. The momentum scale separating the bulk medium from minijets, denoted by $\pjm$, was treated as a free parameter. The framework successfully reproduced soft hadronic observables over a range of $\pjm$ values by readjusting the initial energy-momentum tensor normalization and the specific shear viscosity. This demonstrated that, although minijets capture essential aspects of the underlying physics of the system, hydrodynamic models possess sufficient parameter degeneracy to describe hadronic data with varying amounts of minijet contributions. This degeneracy arises because hadronic observables are primarily sensitive to the late stages of the system's evolution, by which time most minijets have already thermalized, while the surviving partons fragment into hadrons.

Electromagnetic probes --  photons and dileptons -- could serve as useful probes to distinguish between scenarios using different values of $\pjm$. Since they are emitted throughout the entire spacetime evolution of the system and seldom undergo interactions after their production, they retain direct information about the medium from which they originate. Consequently, they are highly sensitive to the dynamical evolution of the plasma.

In this work, we employ the minijet-modified medium as the hydrodynamic background for calculating electromagnetic observables. Our objective is to investigate whether electromagnetic probes are sufficiently sensitive to distinguish between different $\pjm$ values. We find that medium modifications induced by minijets alter both the photon and dilepton spectra, as well as their elliptic flow coefficients $v_2$. In addition, the effective temperature extracted from fits to the dilepton spectrum in the intermediate mass region is also modified. 

The paper is organized as follows. Section~\ref{sec:framework} describes the framework. Section~\ref{sec:hadronic} presents the effect of minijets on hadronic observables and the recalibration of the specific shear viscosity and the normalization of the initial conditions. Section~\ref{sec:em_observables} presents the photon and dilepton observables, including the dilepton's effective temperature. Section~\ref{sec:spacetime} studies the spacetime origin of the electromagnetic sensitivity through a decomposition of the emission into temperature and proper time. Section~\ref{sec:discussion} discusses the results and identifies future extensions including jet-medium photons, heavy quark observables, and dilepton polarization. Throughout this work we use natural units $\hbar = c = 1$ and the mostly minus metric $g^{\mu\nu} = \mathrm{diag}(+1,-1,-1,-1)$.

\section{Framework}\label{sec:framework}
\begin{figure}[t]
\centering
\begin{tikzpicture}[
  font=\scriptsize, >=Latex, node distance=1.2cm,
  box/.style={draw=accent, thick, rounded corners=3pt, fill=accent!8,
              minimum width=2.5cm, minimum height=0.9cm, align=center},
  newbox/.style={draw=newred, very thick, rounded corners=3pt, fill=newred!7,
              minimum width=2.1cm, minimum height=0.9cm, align=center},
  a/.style={->, thick, accent},
  ra/.style={->, line width=1.1pt, newred}
]
\node[box] (music) {\textbf{MUSIC+sources}\\\scriptsize viscous\\hydrodynamics};
\node[box, left=0.9cm of music]  (ipg)   {\textbf{IP-Glasma}\\\scriptsize initial state};
\node[box, right=0.9cm of music] (urqmd) {\textbf{UrQMD}\\\scriptsize hadronic transport};
\node[box, above=0.9cm of ipg]   (pythia){\textbf{PYTHIA}\\\scriptsize minijet production};
\node[box, above=0.9cm of music] (hybrid){\textbf{Hybrid Model}\\\scriptsize parton evolution };
\node[newbox, below=0.9cm of music] (em)
  {\textbf{Thermal Photons}\\$+$\\ \textbf{QGP Dileptons}};
\node[newbox, below=0.9cm of ipg] (prompt) {\textbf{pQCD photons}};
\draw[a] (ipg) -- (music);
\draw[a] (ipg) -- node[midway, right, align=center] {position of\\binary\\collisions} (pythia);
\draw[a] (music) -- (urqmd);
\draw[a] (pythia) -- (hybrid);
\draw[<->, thick, accent] (hybrid) -- (music);
\draw[a] (hybrid.east) -- ($(urqmd.north)+(0,0.0)$);
\draw[ra] (music) -- (em);
\draw[ra] (ipg) -- node[midway, right, align=center] {number of\\binary\\collisions} (prompt);
\end{tikzpicture}
\caption{Schematic of the multistage framework used in our work.}
\label{fig:framework}
\end{figure}
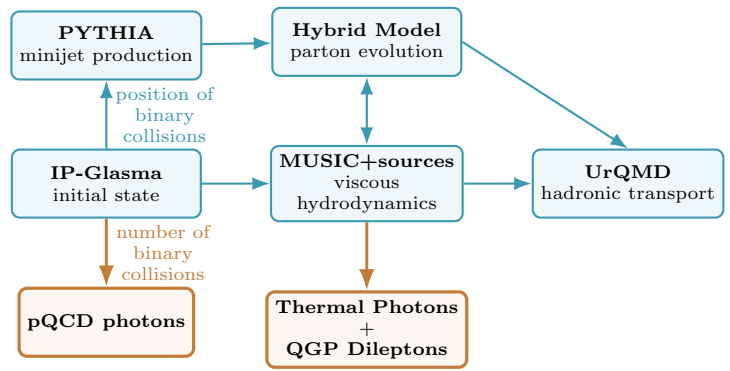

Our approach, shown schematically in Figure~\ref{fig:framework}, is built around a concurrent minijet-hydrodynamics evolution~\cite{Pablos:2022piv}, which provides the background to calculate electromagnetic probes. The initial energy-momentum distribution is generated with IP-Glasma~\cite{Schenke:2012hg,Schenke:2012wb} and the accompanying population of semihard minijets by PYTHIA\,8~\cite{Sjostrand:2006za,Sjostrand:2007gs}. The bulk medium is evolved with relativistic viscous hydrodynamics using MUSIC~\cite{Schenke:2010rr,Schenke:2010nt,Paquet:2015lta}, sourced concurrently by the minijets deposit as they lose energy and momentum in the plasma as described by the Hybrid Model~\cite{Casalderrey-Solana:2014bpa,Casalderrey-Solana:2015vaa}. This is followed by hadronic rescattering in UrQMD~\cite{Bass:1998ca,Bleicher:1999xi}, which receives contributions from both the Cooper-Frye particlization of the plasma and the fragmentation of the minijet partons that did not fully thermalize. Thermal photons and dileptons are computed on this background using the formalism from \cite{Gale:2021emg,Gao:2026vxs}. Prompt photon contributions are computed with pQCD. We describe each stage in turn.

\subsection{Initial state and minijet production}\label{sec:ipglasma}

Event-by-event initial conditions for Pb$+$Pb collisions at $\sqrt{s_{\rm NN}}=2.76$~TeV are generated with IP-Glasma. The model builds on the impact parameter dependent saturation model (IP-Sat) \cite{Kowalski:2003hm}, which is fitted to HERA deep inelastic scattering data \cite{Rezaeian:2012ji} and determines the gluon saturation scale $Q_s$. Within the Color Glass Condensate (CGC)~\cite{Gelis:2010nm} effective field theory, the classical Yang-Mills equations are then solved numerically to obtain the initial gluon fields and the energy-momentum tensor $T^{\mu\nu}$ at the hydrodynamic initialization time. 

The locations of all binary collisions are recorded by IP-Glasma during the event initialization. At each such location, a hard QCD scattering is generated with PYTHIA\,8 using a minimum transverse momentum scale $\pjm$, and the candidate minijet pair is accepted with probability $\sigma_{\mathrm{hard}}/\sigma_{\rm NN}$, where $\sigma_{\mathrm{hard}}$ is the cross section of the generated process and $\sigma_{\rm NN}=64$~mb~\cite{Nakamura:2010zzi} is the total inelastic nucleon-nucleon cross section at $\sqrt{s_{\rm NN}}=2.76$~TeV. Since parton distributions inside a nucleus differ from those in a free proton, nuclear modifications to the parton distributions are incorporated at leading order through EPS09~\cite{Eskola:2009uj}.
 
The threshold $\pjm$ corresponds to the minimum transverse momentum ($\hat{p}_{T,{\rm min}}$) that the hard processes in PYTHIA are allowed to generate, and is the central control parameter in this study. Lowering it increases the number and total energy of the produced minijets, thereby increasing the energy deposited into the hydrodynamic medium. All values of $\pjm$ considered in this work satisfy $\pjm > Q_s$, reducing the overlap between the semihard minijet production described by PYTHIA\,8 and the soft dynamics already captured by the initial state in IP-Glasma. Without this separation, including minijet processes down to the saturation scale would double count degrees of freedom. Some overlap is nevertheless unavoidable, as discussed in Sec.~\ref{sec:hadronic} where we will see that the extreme case for $\pjm=4$~GeV likely overcounts semihard partons. We consider $\pjm=10,7,6,5,4$~GeV together with a no-jets baseline in which the source term is absent, yielding six scenarios whose consequences are studied through the remainder of the paper.

\subsection{Hydrodynamic evolution and the source term}\label{sec:hydro}

The soft sector generated above is evolved with MUSIC, a $3\!+\!1$D viscous relativistic hydrodynamics code implementing the second-order DNMR formalism~\cite{Denicol:2012cn,Molnar:2013lta}. To account for the energy and momentum that the minijets deposit into the medium, the conservation equations carry a source term,

\begin{equation}\label{eq:cons}
\nabla_\mu T^{\mu\nu}_{\mathrm{hydro}} = J^\nu ,
\end{equation}
with the energy-momentum tensor,

\begin{equation}
T^{\mu\nu}_{\mathrm{hydro}} = \epsilon \,u^\mu u^\nu - (P+\Pi)\, \Delta^{\mu\nu} + \pi^{\mu\nu},
\end{equation}
where $\epsilon$ is the energy density, $P$ the equilibrium pressure, $u^\mu$ the fluid four-velocity, $\Pi$ the bulk viscous pressure, $\pi^{\mu\nu}$ the shear stress tensor, and $\Delta^{\mu\nu}=g^{\mu\nu}-u^\mu u^\nu$ the projector orthogonal to $u^\mu$. The equation of state is taken by smoothly matching lattice QCD calculation~\cite{HotQCD:2014kol} at high temperatures to hadronic gas equation of state at low temperatures.

The source term representing minijet energy-momentum deposition is smeared with a Gaussian profile of transverse width $\sigma_T$ and rapidity width $\sigma_\eta$ in Milne coordinates,
\begin{equation}\label{eq:source}
J^\nu(x) = \sum_i   \frac{\Delta P_i^\nu}{\Delta\tau\,(2\pi)^{3/2}\,\sigma_T^2\,\sigma_\eta\,\tau}   \exp\!\left[-\frac{\Delta x_i^2+\Delta y_i^2}{2\sigma_T^2}-\frac{\Delta\eta_i^2}{2\sigma_\eta^2}\right],
\end{equation}
where the sum runs over all four-momentum depositions $\Delta P_i^\nu$ occurring during the hydrodynamic time step $\Delta\tau$, and $\Delta x_i$, $\Delta y_i$, $\Delta\eta_i$ denote the separations between each deposition point and the fluid cell at which the source is evaluated. The Gaussian widths are set to $\sigma_T = 0.4/\sqrt{2}$~fm and $\sigma_\eta = 0.4/\sqrt{2}$~\cite{Pablos:2022piv}.

The shear viscosity over entropy density, $\eta/s$, is taken to be temperature independent and is retuned for each $\pjm$ scenario, as described in Sec.~\ref{sec:hadronic}. The specific bulk viscosity is parametrized~\cite{Schenke:2020mbo} as a temperature dependent asymmetric Gaussian peaked near the crossover
region,

\begin{equation}\label{eq:bulk}
\frac{\zeta}{s}(T) =
\begin{cases}
B_{\rm norm}\,\exp\!\left[-\dfrac{(T-T_{\rm peak})^2}{2\,B_1^2}\right],
  & T < T_{\rm peak},\\[8pt]
B_{\rm norm}\,\exp\!\left[-\dfrac{(T-T_{\rm peak})^2}{2\,B_2^2}\right],
  & T > T_{\rm peak},
\end{cases}
\end{equation}
with $B_{\rm norm}=0.13$, $B_1=0.01$~GeV, $B_2=0.12$~GeV, and $T_{\rm peak}=0.16$~GeV. These bulk viscosity parameters are held fixed across all scenarios.

\subsection{Concurrent minijet-hydrodynamics evolution}\label{sec:concurrent}

The minijet and hydrodynamic evolutions proceed concurrently: at each hydrodynamic time step, the positions and momenta of all minijet partons are updated, and the energy-momentum they lose is deposited into the medium through the source term $J^\nu$ in Eq.~\eqref{eq:cons}. The medium, in turn, evolves in response to this deposition, so that the local temperature and flow velocity seen by each parton reflect the accumulated effect of all prior depositions.

The in-medium energy loss of each parton is modeled within the hybrid strong/weak coupling framework~\cite{Casalderrey-Solana:2014bpa,Casalderrey-Solana:2015vaa}. High virtuality splittings ($Q^2\gg T^2$) are treated perturbatively using vacuum DGLAP evolution, while the soft momentum exchange between each parton and the medium ($q\sim T$) is described by an energy-loss rate inspired by holography~\cite{Chesler:2014jva,Chesler:2015nqz},

\begin{equation}\label{eq:dEdx}
\frac{1}{E_{\rm in}}\frac{dE}{dx} = -\frac{4}{\pi} \frac{x^2}{x^2_{\rm stop}} \frac{1}{\sqrt{x_{\rm stop}^{2}-x^{2}}}\,,
\end{equation}
where $E_{\rm in}$ is the energy of the parton at the start of its in-medium propagation and $x_{\rm stop}$~\cite{Chesler:2008uy,Gubser:2008as} is the stopping distance over which it deposits its entire energy,

\begin{equation}\label{eq:xstop}
x_{\rm stop} = \frac{1}{\kappa\, T} \left(\frac{E_{\rm in}}{T}\right)^{1/3}.
\end{equation}

The dimensionless coefficient $\kappa$ is species dependent: for quarks we take $\kappa_q = 0.4$, as extracted from phenomenological studies of jet quenching at strong coupling~\cite{Casalderrey-Solana:2018wrw}, while for gluons $\kappa_g = (C_A/C_F)^{1/3}\,\kappa_q$~\cite{Gubser:2008as}, where $C_A$ and $C_F$ are the Casimir invariants of the adjoint and fundamental representations of SU(3), reflecting the shorter stopping distance of a gluon.
Rarer, more energetic momentum exchanges with $q\gg m_D$ (where $m_D$ is the Debye screening mass) can of course also occur, corresponding to both elastic and inelastic processes which are in the perturbative regime. While some of these processes are now included in the Hybrid Model~\cite{Hulcher:2026dht}, the contribution of these perturbative processes at the energy scales relevant for minijet evolution will be, in this model, relatively small, and have been excluded in the present study.

The hydrodynamic evolution is initialized at $\tau_0 = 0.4$~fm, when the IP-Glasma energy-momentum tensor is matched to the hydrodynamic one including $\pi^{\mu\nu}$ and $\Pi$.
Minijet partons propagate according to a spacetime picture based on formation time arguments~\cite{Casalderrey-Solana:2011fza}, with each parton traveling for its formation time $\tau_f = 2E/Q^2$ before splitting, where $E$ is the parton energy and $Q$ its virtuality.
Before $\tau_0$, no energy loss occurs, ignoring any possible quenching effects during this pre-hydrodynamic stage \cite{Aurenche:2012qk,Andres:2019eus,Ipp:2020nfu,Carrington:2022bnv,Boguslavski:2024ezg}. Energy and momentum deposition through the source term of Eq.~\eqref{eq:source} thus begins at $\tau_0$. Energy loss does not occur for partons located in fluid cells below the switching temperature of $145$~MeV,  even though the resulting hadrons still rescatter in the hadron resonance gas phase through UrQMD (Sec.~\ref{sec:urqmd}).

\subsection{Particlization and hadronic transport}\label{sec:urqmd}

When the local temperature in a fluid cell drops below a switching value of $T_{\mathrm{sw}}=145$~MeV, hadrons are produced via the Cooper-Frye prescription~\cite{cooper1974single}. The freeze-out hypersurface is constructed using the Cornelius algorithm~\cite{Huovinen:2012is}, and hadron sampling is performed with iSS~\cite{Shen:2014vra}.

Minijet partons that have not fully hydrodynamized by the time the surrounding fluid cell reaches $T_{\mathrm{sw}}$ are hadronized independently. Their remaining energy and momentum are fragmented using the PYTHIA\,8 Lund string fragmentation model. Color neutralization of these surviving partons is handled following the procedure described in~\cite{Pablos:2022piv}.
The resulting hadrons are passed to the hadronic afterburner alongside the Cooper-Frye output.

The hadronic rescattering and resonance decays are modeled with UrQMD. These hadrons scatter through elastic and inelastic collisions until all interactions cease. The final-state hadron momenta after UrQMD evolution are used to compute the charged particle multiplicity and elliptic flow that enter the hadronic calibration of Sec.~\ref{sec:hadronic}. For each scenario, $N_{\mathrm{ev}} = 500$ hydrodynamic events are generated per 10\% centrality bin, and each freeze-out hypersurface is oversampled $100$ times in iSS.

\subsection{Direct photons and dilepton emission}\label{sec:emission}

The evolving medium described above radiates electromagnetically at every stage of its history. For the purpose of computing thermal emission, the hydrodynamic evolution is continued past the hadronic switching temperature $T_{\mathrm{sw}}=145$~MeV down to $T=105$~MeV, following~\cite{Paquet:2015lta}, so that late stage photon production in the hadronic phase can be estimated within the fluid dynamic framework. Thermal photon and dilepton yields are obtained by integrating the local emission rates over the full spacetime history of the plasma. For photons, the invariant spectrum is,

\begin{equation}\label{eq:photon_yield}
E\frac{dN_\gamma}{d^3p} = \int d^4x\;  E\frac{dR_\gamma}{d^3p}\!\bigl(T(x),u^\mu(x),\pi^{\mu\nu}(x),\Pi(x)\bigr),
\end{equation}
where $dR_\gamma/d^3p$ is the local emission rate evaluated in each fluid cell at temperature $T$, flow velocity $u^\mu$, shear tensor $\pi^{\mu\nu}$ and bulk pressure $\Pi$. In the QGP phase, the rate includes the complete leading order contribution, $2\!\to\!2$ Compton and annihilation channels~\cite{McLerran:1984ay,Kapusta:1991qp} together with near collinear bremsstrahlung and pair-annihilation processes resummed via the Arnold-Moore-Yaffe (AMY) formalism~\cite{Arnold:2001ms}. In the hadronic phase, the rates comprise the mesonic and baryonic contributions included in~\cite{Turbide:2003si}, supplemented by $\pi\pi$ bremsstrahlung~\cite{Heffernan:2014mla} and the $\pi\rho\to\omega\gamma$, $\pi\omega\to\rho\gamma$, and $\pi\omega\to\rho\pi$ channels~\cite{Holt:2015cda}. The QGP and hadronic rates are switched over a narrow temperature interval around $T \approx 180$~MeV. Both thermal and prompt photon spectra are evaluated at midrapidity ($y=0$). Where available, the thermal photon rates \cite{photonem} include the shear and bulk viscous corrections to the distribution functions, following~\cite{Paquet:2015lta}.

Prompt photons are computed at next-to-leading order in perturbative QCD using INCNLO~\cite{INCNLO}, with CTEQ61m~\cite{Stump:2003yu} proton parton distribution functions, EPS09~\cite{Eskola:2009uj} nuclear modifications, and BFG-2~\cite{Bourhis:1997yu} parton to photon fragmentation functions. The factorization, renormalization, and fragmentation scales are all set equal, $\mu_F=\mu_R=\mu_{\rm frag}=p_T/2$. Because parton distribution and fragmentation functions are only defined above a minimum scale $Q_0\approx1.5$~GeV, the calculation cannot be evaluated directly at $\mu=p_T/2$ for low $p_T$. Since varying the scale changes primarily the normalization of the prompt spectrum rather than its $p_T$ dependence, the spectrum is extended to low $p_T$ by evaluating the calculation at a larger scale and renormalizing it to match the $\mu=p_T/2$ result at high $p_T$, following the procedure of~\cite{Paquet:2015lta,Paquet:2016ulk}. The prompt yield in Pb$+$Pb collisions is obtained by scaling the nucleon-nucleon result by the number of binary collisions in each event.

For dileptons, the invariant mass spectrum is obtained similarly,

\begin{equation}\label{eq:dilepton_yield}
\frac{dN_{e^+e^-}}{dM} = \int d^4x \frac{dR_{e^+e^-}}{dM}\!\bigl(T(x),u^\mu(x)\bigr),
\end{equation}
where the integration runs over the QGP phase of the evolution, in fluid cells  with $T > 180$~MeV. We have not included hadronic dilepton sources. Their contribution is small in the intermediate mass region considered here, where partonic emission dominates~\cite{Gao:2026vxs}.
The QGP dilepton emission rate is evaluated~\cite{Gao:2026vxs} using the vector channel spectral function $\rho_V = \rho^\mu{}_\mu$ at next-to-leading order in the strong coupling. At leading order, the rate is given by the Born process $q\bar{q}\to\gamma^*\to e^+e^-$. The NLO corrections include Compton scattering ($gq\to\gamma^* q$, $g\bar{q}\to\gamma^*\bar{q}$), modified annihilation ($q\bar{q}\to\gamma^* g$), one loop virtual corrections~\cite{Jackson:2019mop,Laine:2013vma} to the Born process, and Landau-Pomeranchuk-Migdal (LPM) resummation~\cite{Ghiglieri:2013gia,Ghiglieri:2014kma,Aurenche:2002pc,Aurenche:2002wq,Arnold:2001ms,Arnold:2001ba}. We label the sum of these contributions LO$+$NLO throughout this work. Viscous corrections to the dilepton emission rates are not included, as these are not yet available at NLO. The dilepton yields are integrated over the rapidity interval $|y|<0.8$. Drell-Yan contributions are not included in the intermediate mass region, where thermal QGP emission dominates the dilepton yield.


\section{Hadronic observables and parameter degeneracy}\label{sec:hadronic}

With the framework of Sec.~\ref{sec:framework} in place, we first confront it with hadronic data. Compared to~\cite{Pablos:2022piv}, in this work we initialize the hydrodynamic energy-momentum tensor by including the dissipative components ($\pi^{\mu \nu}$ and $\Pi$) and we use the parameter set of~\cite{Schenke:2020mbo} that describes broad range of observables across a wide range of collision systems. The minijet source term injects additional energy and momentum into the hydrodynamic medium. The added gradients cause additional entropy production, with the total amount increasing as the threshold $\pjm$ is lowered. 
To identify the effect of minijets on electromagnetic observables, each scenario must first be calibrated with respect to hadronic data. For each value of $\pjm$, we retune two parameters: (i) the overall initial normalization factor  which rescales the initial energy-momentum tensor from IP-Glasma at the hydrodynamics matching time $\tau_0$, and (ii) the shear viscosity over entropy density $\eta/s$.  The normalization is adjusted to reproduce the centrality-dependent charged particle multiplicity $dN_{\mathrm{ch}}/d\eta$ measured by ALICE~\cite{ALICE:2010mlf}, and $\eta/s$ is then tuned to match the charged hadron elliptic flow $v_2$~\cite{ALICE:2011ab} in the same centrality bins. Both parameters decrease with decreasing $\pjm$ because more minijets require a smaller initial normalization to preserve the final state multiplicity, and a smaller $\eta/s$ to compensate for the altered gradients caused by randomly oriented minijet deposition. The resulting parameter sets are collected in Table~\ref{tab:params}. While minijets may contribute only a small fraction of the system’s total energy-momentum, the additional gradients can cause substantial entropy production, necessitating a significant recalibration of $s_{\mathrm{factor}}$.

We emphasize that $\pjm = 4$ GeV represents an extreme case in which intermediate $p_T$ partons are likely doublecounted, as also observed in \cite{Pablos:2022piv}. A simple rescaling of $s_{\mathrm{factor}}$ does not selectively remove modes already accounted for by the minijets, leading to doublecounting. Nevertheless, we include this case to provide an upper bound on the effect of minijets. \footnote{A recent study \cite{Faraday:2026ztw} simultaneously produces soft and semi-hard modes from Glasma fields and could be incorporated into our setup to avoid double counting.}

\begin{table}[t]
\centering
\caption{Initial energy-momentum normalization $s_{\mathrm{factor}}$ and shear viscosity over entropy density $\eta/s$ for each minijet scenario in Pb$+$Pb collisions at $\sqrt{s_{\rm NN}}=2.76$~TeV. Both parameters decrease monotonically with decreasing $\pjm$.}
\label{tab:params}
\begin{ruledtabular}
\begin{tabular}{lcc}
Scenario & $s_{\mathrm{factor}}$ & $\eta/s$ \\
\hline
No jets          & 0.260 & 0.135 \\
$\pjm=10$~GeV    & 0.244 & 0.125 \\
$\pjm= 7$~GeV    & 0.224 & 0.100 \\
$\pjm= 6$~GeV    & 0.200 & 0.095 \\
$\pjm= 5$~GeV    & 0.170 & 0.085 \\
$\pjm= 4$~GeV    & 0.120 & 0.055 \\
\end{tabular}
\end{ruledtabular}
\end{table}

\begin{figure}[t]
\centering
\includegraphics[width=\columnwidth]{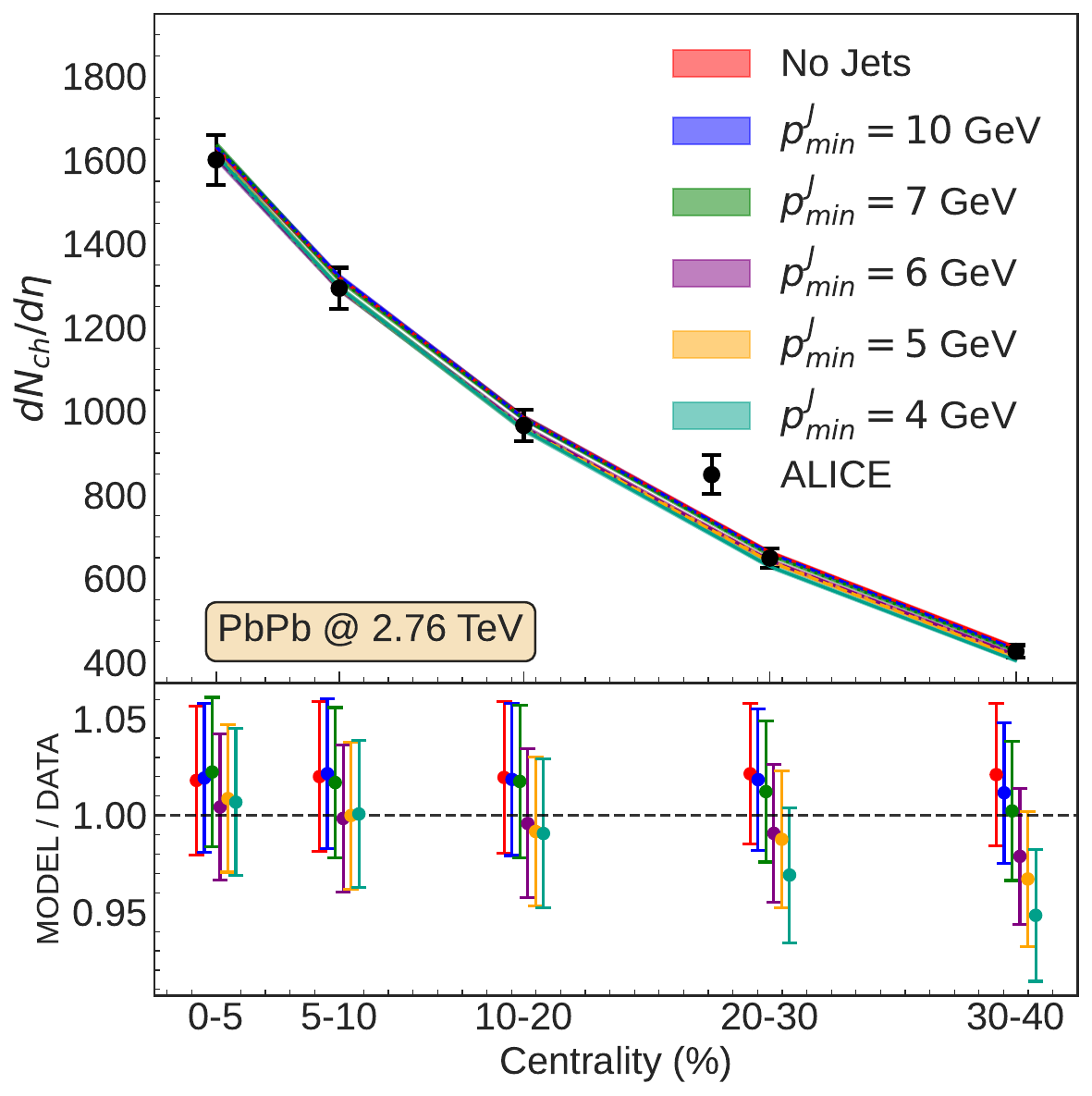}
\caption{Charged particle multiplicity $dN_{\rm ch}/d\eta$ versus centrality in Pb$+$Pb collisions at $\sqrt{s_{NN}}=2.76$~TeV, for the no-jets
baseline and for $\pjm=10, 7, 6, 5, 4$~GeV, after retuning the initial normalization $s_{\mathrm{factor}}$ and $\eta/s$ in each scenario, compared to ALICE data~\cite{ALICE:2010mlf}.}
\label{fig:multiplicity}
\end{figure}

\begin{figure}[t]
\centering
\includegraphics[width=\columnwidth]{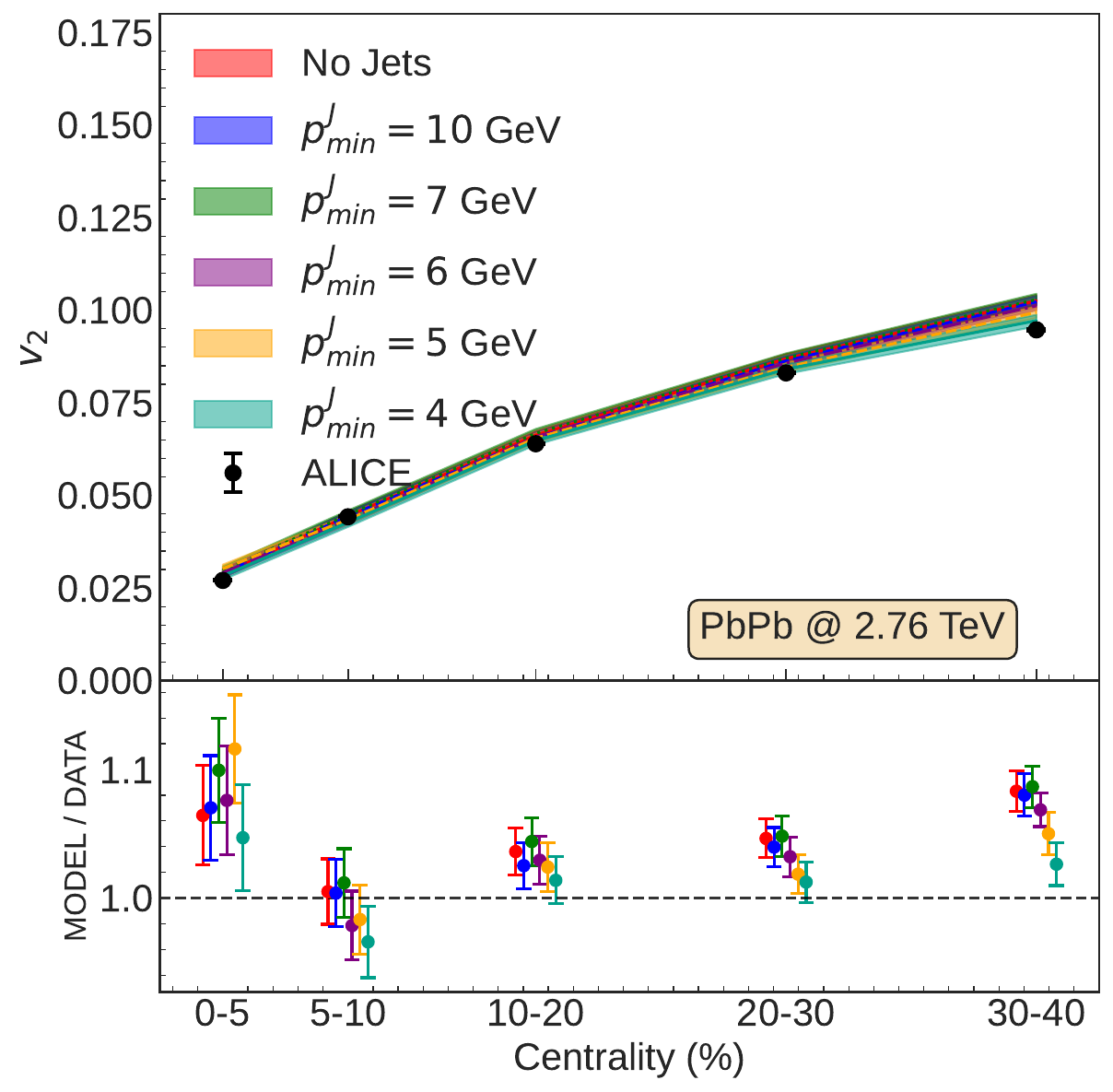}
\caption{Charged hadron elliptic flow $v_2$ versus centrality for the same six scenarios as in Fig.~\ref{fig:multiplicity}, compared to ALICE data~\cite{ALICE:2011ab}.}
\label{fig:hadronv2}
\end{figure}

\begin{figure}[t]
\centering
\includegraphics[width=\columnwidth]{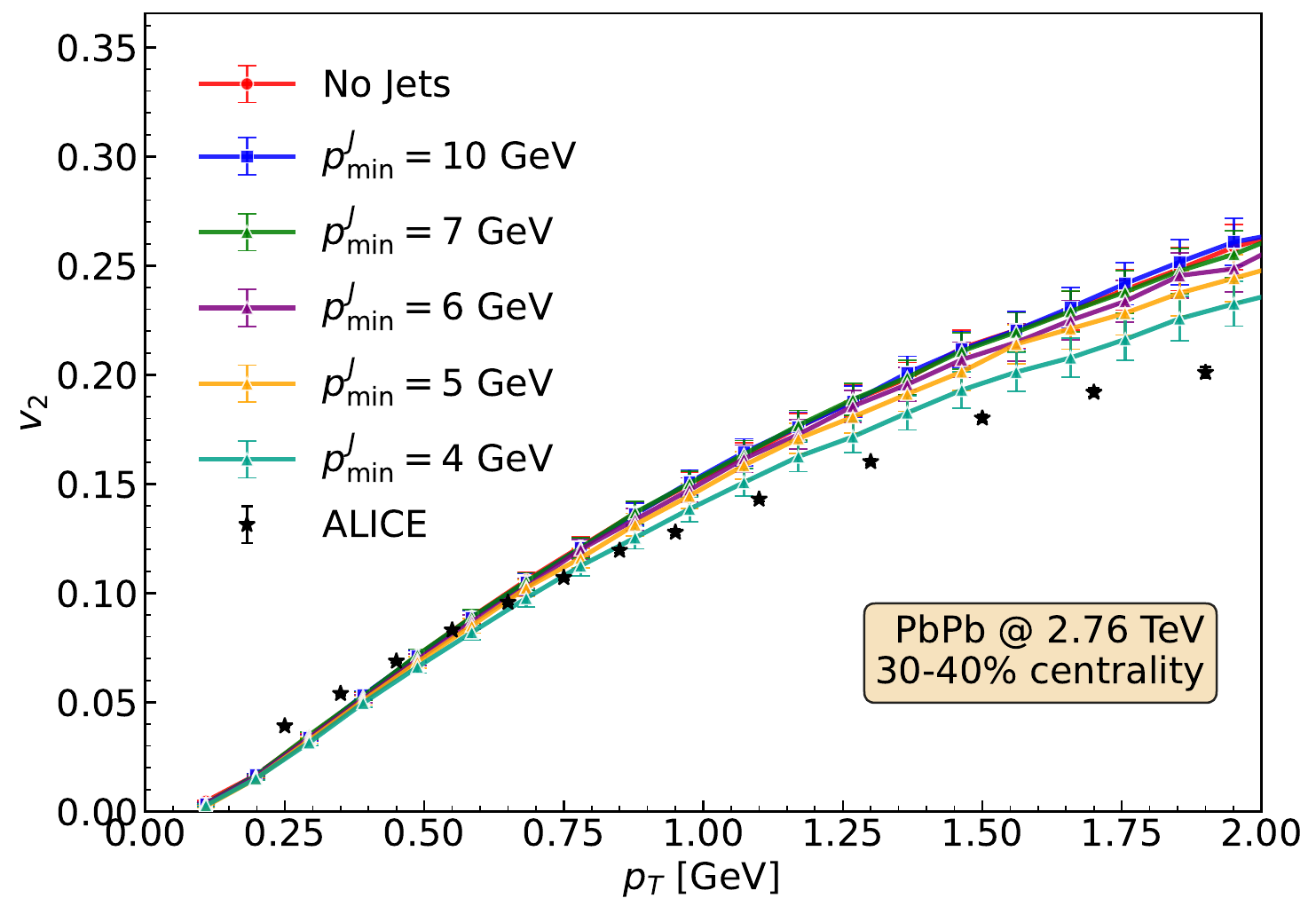}
\caption{Charged hadron elliptic flow $v_2\{\mathrm{SP}\}$ versus $p_T$ in 30-40\% central Pb$+$Pb collisions at $\sqrt{s_{NN}}=2.76$~TeV, for the same six scenarios as in Fig.~\ref{fig:multiplicity}, compared to ALICE data~\cite{ALICE:2011ab}. Uncertainties on the theoretical curves are statistical.}
\label{fig:hadronv2pT_c34}
\end{figure}

\begin{figure}[t]
\centering
\includegraphics[width=\columnwidth]{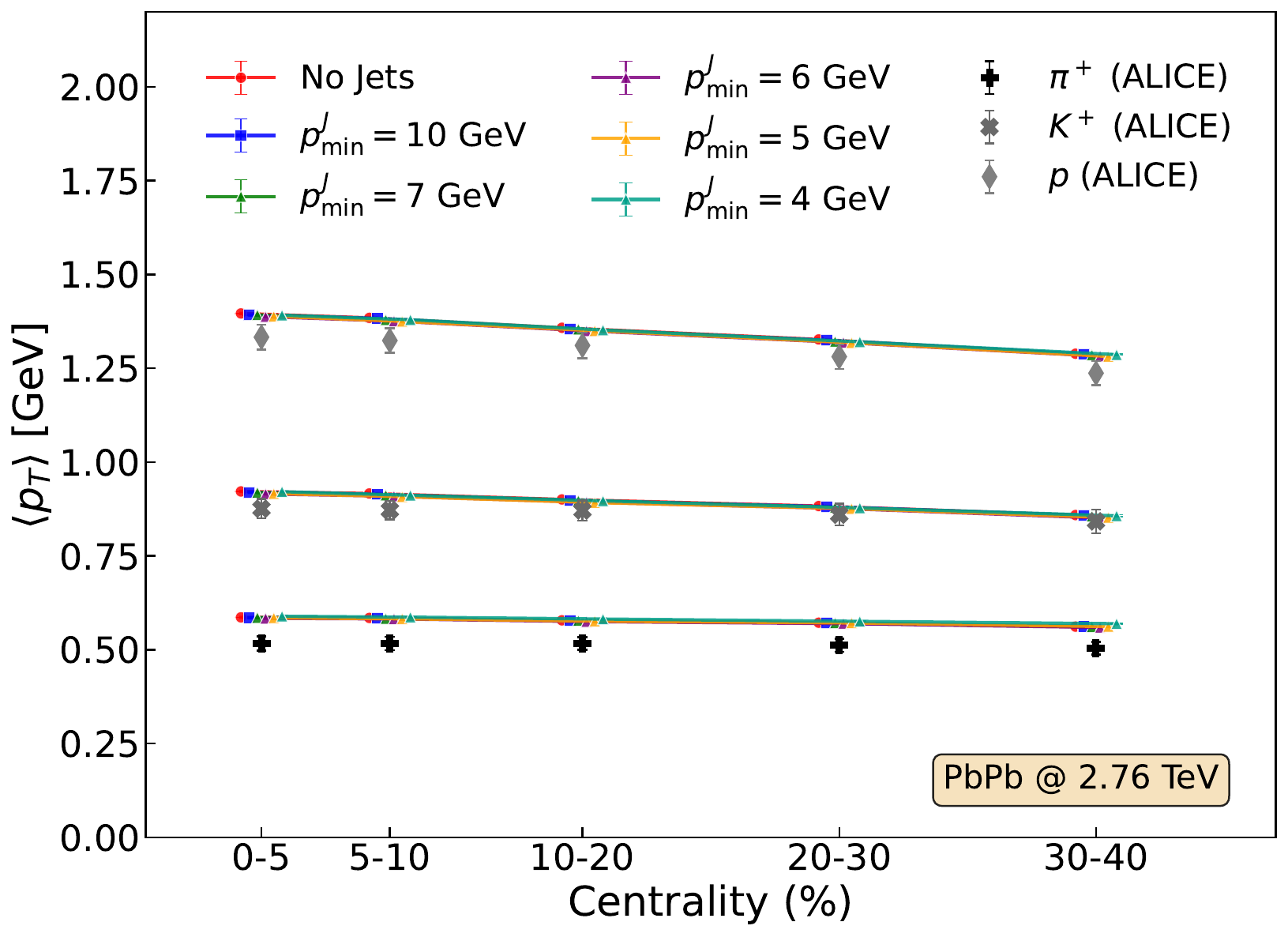}
\caption{Mean transverse momentum $\langle p_T \rangle$ of $\pi^+$, $K^+$, and $p$ versus centrality, for the same six scenarios as in Fig.~\ref{fig:multiplicity}, compared to ALICE data~\cite{ALICE:2013mez}. Uncertainties on the theoretical curves are statistical; those on the data combine statistical and systematic uncertainties.}
\label{fig:meanpT_pos}
\end{figure}

Figures~\ref{fig:multiplicity} and~\ref{fig:hadronv2} demonstrate the resulting degeneracy. All six scenarios reproduce the measured $dN_{\mathrm{ch}}/d\eta$ across centralities (Fig.~\ref{fig:multiplicity}), and the charged hadron $v_2$ is likewise indistinguishable among them (Fig.~\ref{fig:hadronv2}), despite a factor of ${\sim}\,2.5$ variation in $\eta/s$ from the no-jets baseline to $\pjm=4$~GeV. This degeneracy extends to the transverse momentum dependence of charged hadron elliptic flow shown in Fig.~\ref{fig:hadronv2pT_c34} in $30-40$\% centrality bin, where the six curves are consistent with each other. Figure~\ref{fig:meanpT_pos} confirm that the mean transverse momentum of identified hadrons ($\pi^+, K^+,p$) is similarly insensitive to the minijet scenarios. The difference in mean $p_T$ among different $\pjm$ is of the order of 1\%, consistent with the observed differences in charged-hadron yields. This is because particle spectra are unchanged once parameters have been retuned \cite{Pablos:2022piv}. 
Even though the two parameters were retuned to fit just the charged hadron multiplicity and integrated charged hadron $v_2$, the resulting setups can explain other hadronic observables as well. For comparison with more hadronic observables, see~\cite{Pablos:2022piv}.

This $\pjm$-$s_{\mathrm{factor}}$-$\eta/s$ degeneracy is the central observation: the hadronic bulk observables considered here cannot determine whether minijets contribute to the medium evolution, nor constrain $\eta/s$ without an independent handle on the initial normalization. This motivates examining electromagnetic probes, which are sensitive to the early, hot stage of the evolution affected by this retuning. In particular, we investigate whether photons and dileptons can distinguish between these otherwise degenerate scenarios. Section~\ref{sec:em_observables} addresses this question.

\section{Electromagnetic observables}\label{sec:em_observables}
\subsection{Photons}

Having seen in the previous Section how limited is the sensitivity of hadronic observables to scenarios with widely different minijet abundances, we now turn to electromagnetic probes. We first look at direct photons, which we compute as the sum of thermal and prompt contributions, and investigate how they can be used to probe the modified spacetime evolution dynamics caused by minijets.

\begin{figure}[t]
\centering
\includegraphics[width=\columnwidth]{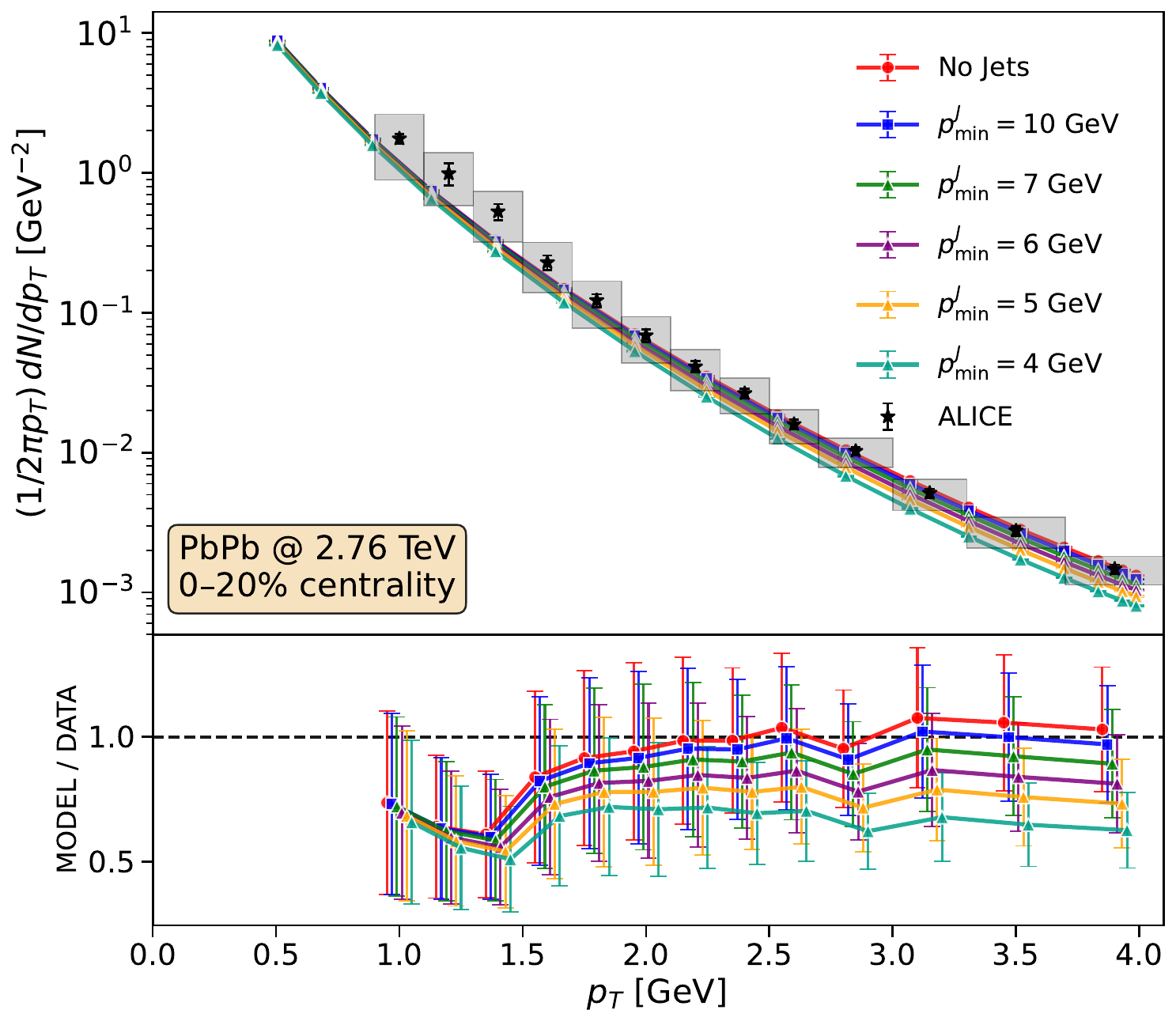}
\caption{Direct photon transverse momentum spectrum in 0-20\% central Pb$+$Pb collisions at $\sqrt{s_{NN}}=2.76$~TeV, for the no-jets baseline and $\pjm=10,7,6,5,4$~GeV, compared to ALICE data~\cite{ALICE:2015xmh}. Uncertainties shown on the theoretical curves are statistical. On the experimental data, the boxes represent systematic and the vertical bars represent statistical uncertainties. The model to data ratio is shown in the lower panel.}
\label{fig:photonspectra020}
\end{figure}

\begin{figure}[t]
\centering
\includegraphics[width=\columnwidth]{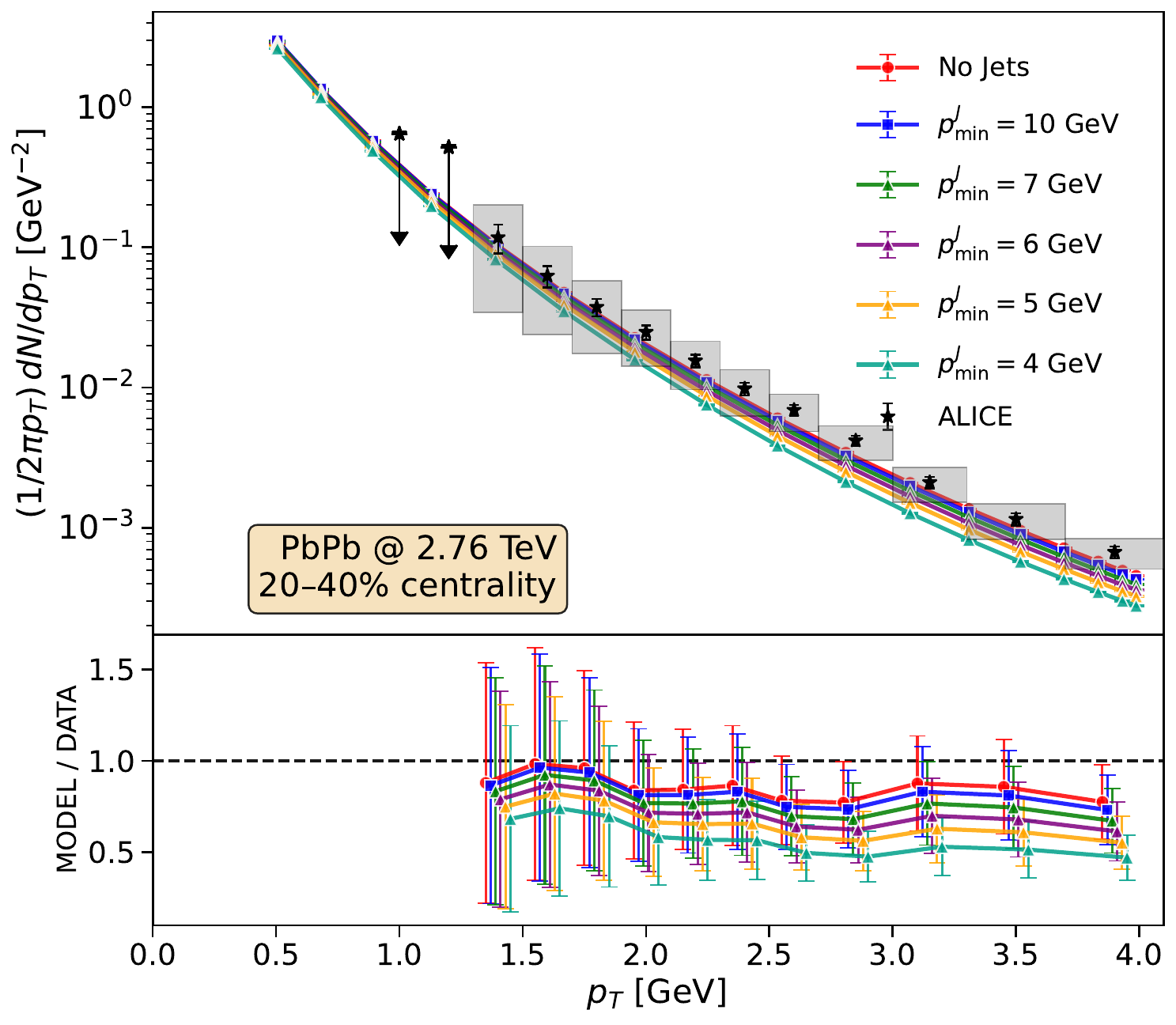}
\caption{Same as Fig.~\ref{fig:photonspectra020}, for 20-40\% centrality.}
\label{fig:photonspectra2040}
\end{figure}

Figures~\ref{fig:photonspectra020} and~\ref{fig:photonspectra2040} show the direct photon transverse momentum spectrum in 0-20\% and 20-40\% central Pb$+$Pb collisions at $\sqrt{s_{NN}}=2.76$~TeV, compared to ALICE data~\cite{ALICE:2015xmh}. A clear ordering emerges among the scenarios: lower $\pjm$ yields a softer direct photon spectrum, with the separation growing toward high $p_T$ where early time dynamics is most prominent. This ordering reflects the reduced initial temperature imposed by the smaller $s_{\mathrm{factor}}$ in the more minijet rich scenarios. At low $p_T$, the spectra converge, as late stage emission, common to all scenarios, dominates.

\begin{figure}[t]
\centering
\includegraphics[width=\columnwidth]{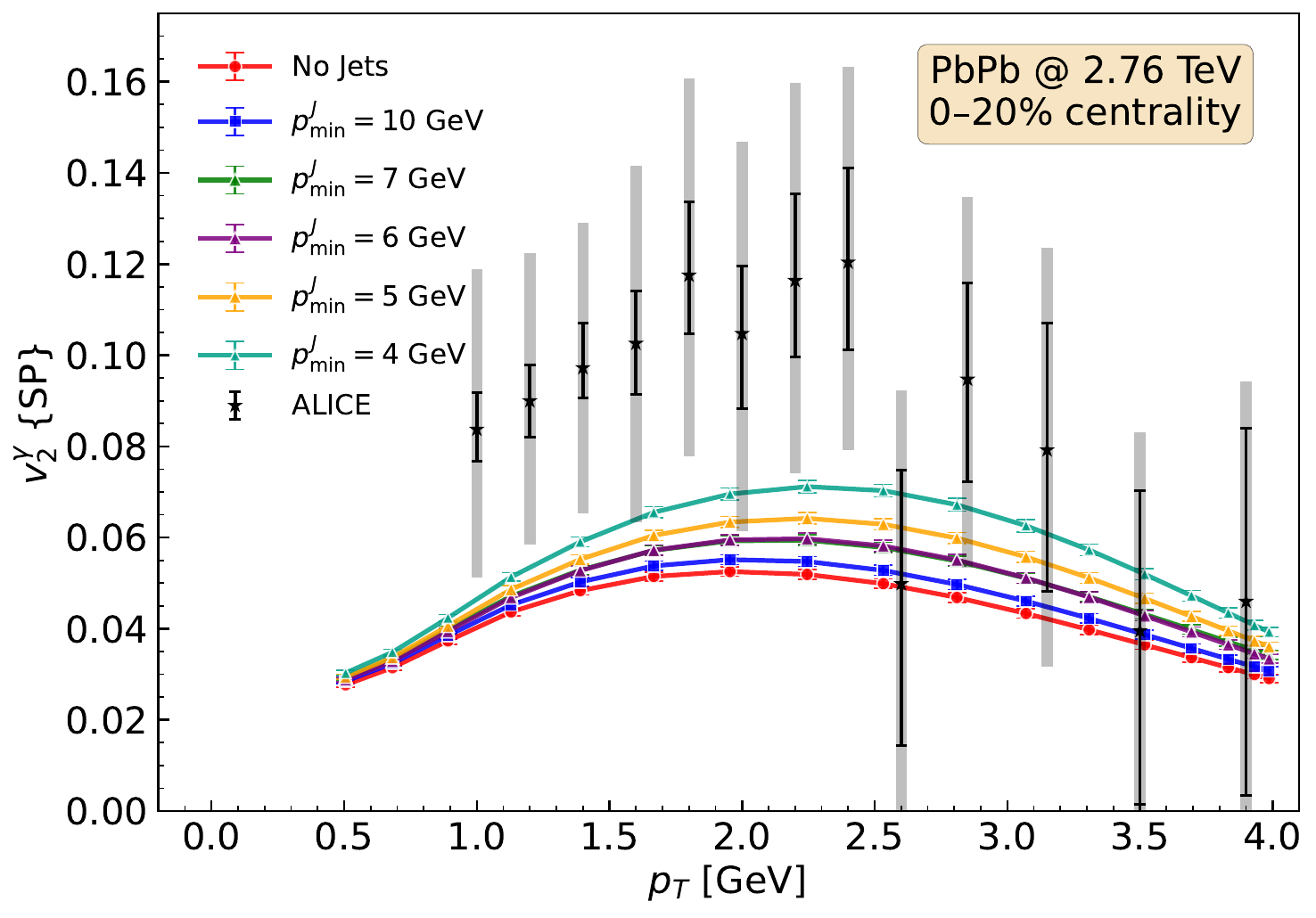}
\caption{Direct photon elliptic flow $v_2^\gamma 
\{\mathrm{SP}\}$ versus $p_T$ in 0-20\% centrality, for the same six scenarios, compared to ALICE data~\cite{ALICE:2018dti}. The ordering inherited from the hadronic tuning propagates into the photon $v_2$: smaller $\eta/s$ (lower $\pjm$) gives larger $v_2$. Uncertainties shown on the theoretical curves are statistical (jackknife). On the experimental data, the boxes represent total and the vertical bars represent statistical uncertainties.}
\label{fig:photonv2020}
\end{figure}

\begin{figure}[t]
\centering
\includegraphics[width=\columnwidth]{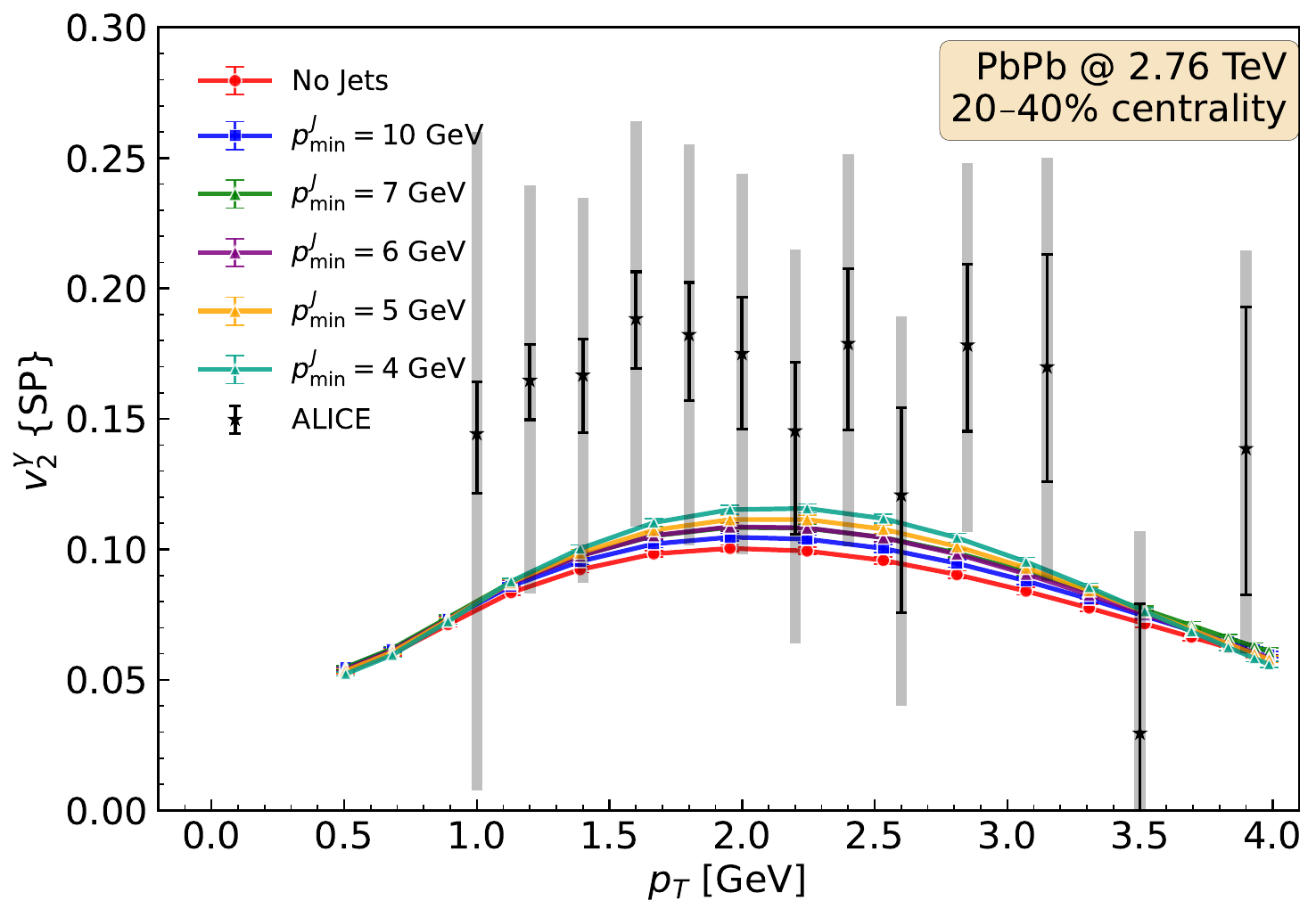}
\caption{Same as Fig.~\ref{fig:photonv2020}, for 20-40\% centrality.}
\label{fig:photonv22040}
\end{figure}

The direct photon elliptic flow $v_2\{\mathrm{SP}\}$ as a function of $p_T$ is shown in Figs.~\ref{fig:photonv2020} and~\ref{fig:photonv22040} for 0-20\% and 20-40\% centrality, respectively. The $v_2$ is computed with the scalar product method~\cite{Lohner:2013blo}.

\begin{equation}\label{eq:v2sp}
v_n^{\gamma}\{\mathrm{SP}\}(p_T^\gamma) = \frac{\bigl\langle v_n^{\gamma}(p_T^\gamma)\;v_n^{h}\; \cos\bigl[n\bigl(\Psi_n^{\gamma}(p_T^\gamma)-\Psi_n^{h}\bigr)\bigr] \bigr\rangle} {\sqrt{\bigl\langle (v_n^{h})^2 \bigr\rangle}} \,,
\end{equation}
where $h$ stands for charged hadrons. Statistical uncertainties on the theoretical curves are estimated using the jackknife method. A clear monotonic ordering emerges: scenarios with lower $\pjm$ exhibit a larger photon $v_2$. Since minijets are produced isotropically, they tend to reduce the momentum anisotropy and naively we would expect a lowering of $v_2$ with addition of minijets. However, accounting for this effect requires retuning the shear viscosity to reproduce the measured hadron $v_2$. Consequently, a lower $\pjm$ requires a smaller $\eta/s$, which in turn leads to a larger photon $v_2$.

\subsection{Dileptons}

Thermal dileptons offer a complementary probe of the early stage dynamics. Unlike photon $p_T$ spectra,
which are Doppler shifted by the collective flow of the emitting cell \cite{Shen:2013vja}, the dilepton invariant mass spectrum, being a Lorentz scalar, reflects the source temperature \cite{Churchill:2023zkk}. In the intermediate mass region (IMR, $1\lesssim M\lesssim 3$~GeV), higher $M$ preferentially samples hotter matter. Thus, the IMR spectrum carries a direct imprint of the early, high temperature phase where the minijets have the strongest effect.

\begin{figure}[t]
\centering
\includegraphics[width=\columnwidth]{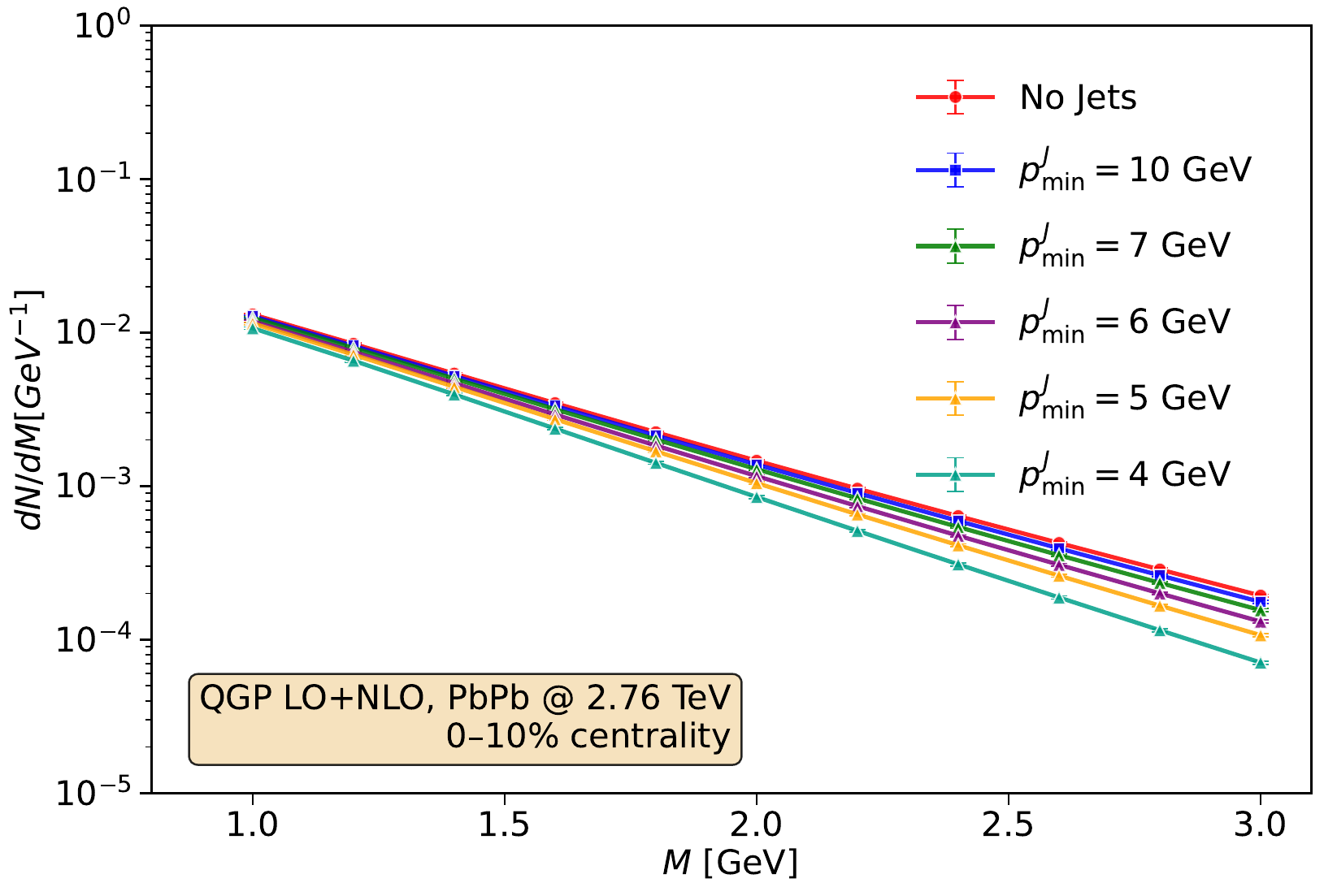}
\caption{Dilepton invariant mass spectrum $dN/dM$ from the QGP LO+NLO channel in 0-10\% central Pb$+$Pb collisions at $\sqrt{s_{NN}}=2.76$~TeV. Uncertainties shown on the theoretical curves are statistical.}
\label{fig:dileptonspectra}
\end{figure}

Figure~\ref{fig:dileptonspectra} shows the dilepton invariant mass spectrum $dN/dM$ from the QGP channel (LO $+$ NLO) in 0-10\% central Pb$+$Pb collisions at $\sqrt{s_{NN}}=2.76$~TeV. In the intermediate mass region, the no-jets baseline produces the highest yield, and the yield decreases monotonically with decreasing $\pjm$. This ordering reflects the same mechanism seen in the photon spectra: lower $\pjm$ requires a smaller $s_{\mathrm{factor}}$, producing a cooler early medium that radiates fewer high-mass pairs.

\begin{figure}[t]
\centering
\includegraphics[width=\columnwidth]{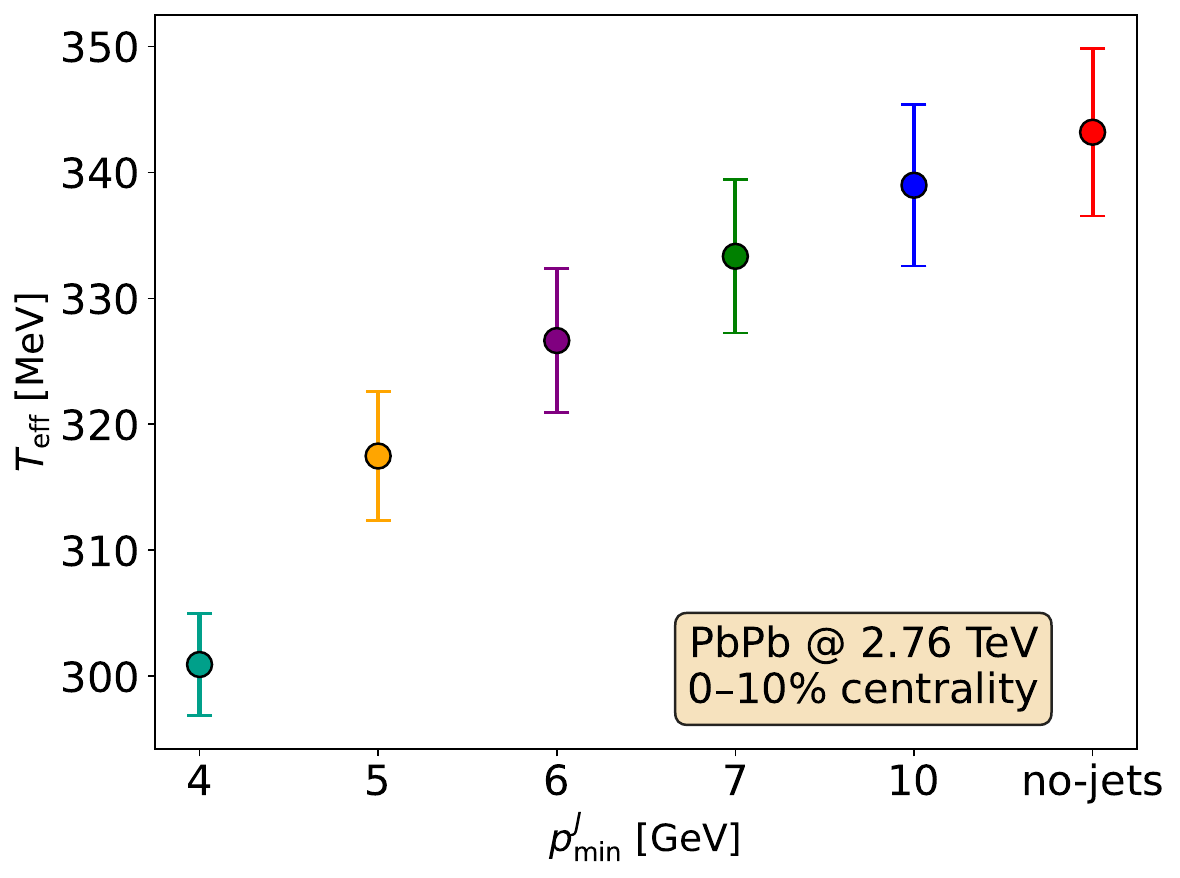}
\caption{Effective temperature $\Teff$ extracted from a fit to the intermediate mass dilepton spectrum in 0-10\% centrality, as a function of $\pjm$.}
\label{fig:Teff}
\end{figure}

We extract an effective temperature $\Teff$ by fitting the invariant mass spectrum to the functional form

\begin{equation}\label{eq:Teff_fit}
\frac{dN}{dM} = A\,M^{3/2}\,e^{-M/\Teff},
\end{equation}
where A is a normalization constant. Figure~\ref{fig:Teff} displays $\Teff$ as a function of $\pjm$. The effective temperature decreases monotonically from $\Teff\approx 345$~MeV for the no-jets baseline to $\Teff\approx 300$~MeV for $\pjm=4$~GeV, a spread of ${\sim}\,45$~MeV. This is a consequence of slow heating as minijets thermalize. Despite the energy injection, the net effect is a cooler medium in the early stages. This is encoded in the IMR dilepton spectrum through $\Teff$.

\begin{figure}[t]
\centering
\includegraphics[width=\columnwidth]{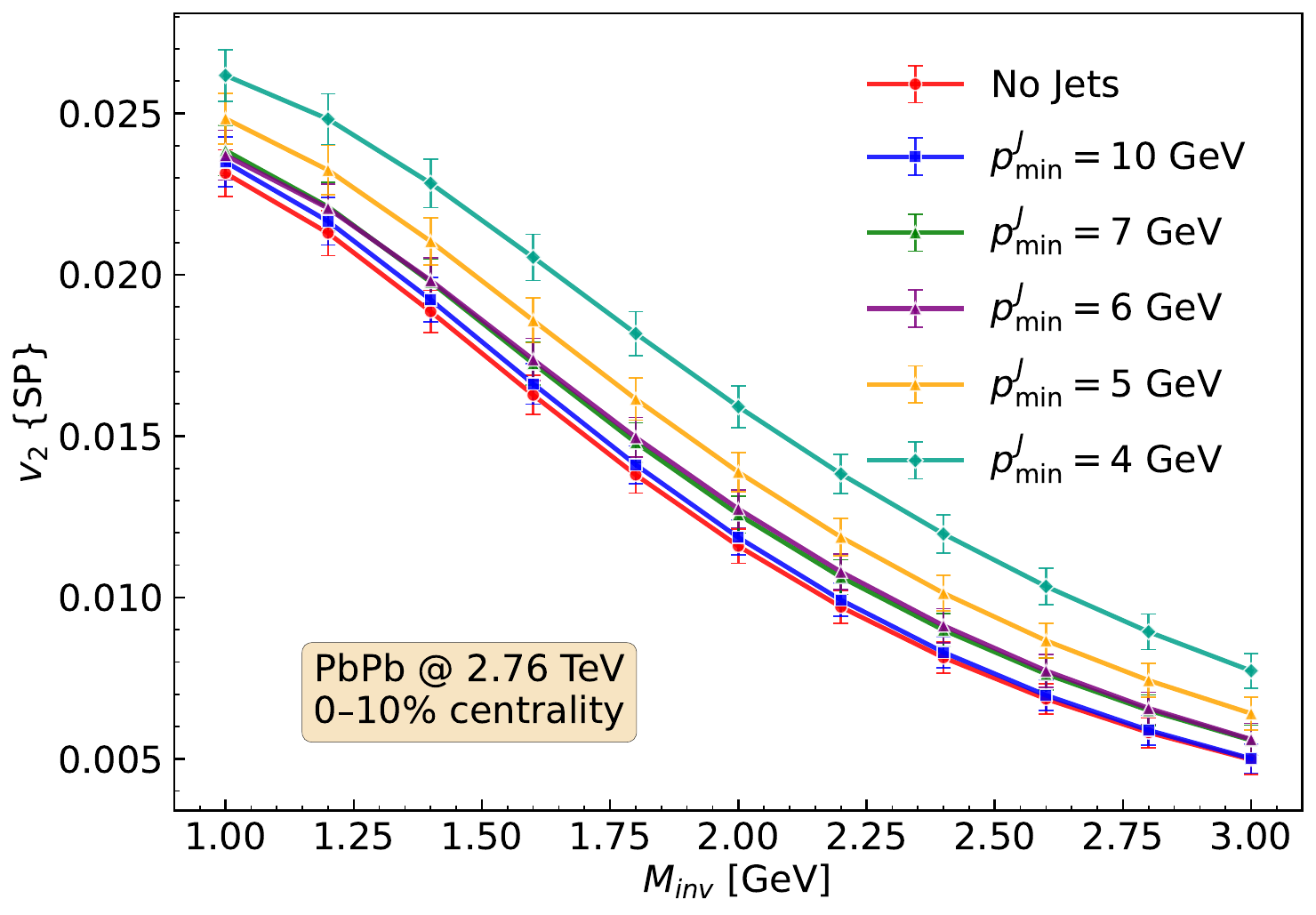}
\caption{Dilepton elliptic flow $v_2\{\mathrm{SP}\}$ versus invariant mass $M$ in 0-10\% centrality, for the same six scenarios.}
\label{fig:dileptonv2}
\end{figure}

Figure~\ref{fig:dileptonv2} shows the dilepton elliptic flow $v_2\{\mathrm{SP}\}$, computed using the scalar product method, as a function of invariant mass $M$ in 0-10\% centrality. The same ordering observed in the direct-photon $v_2$ carries over: lower $\pjm$ (smaller $\eta/s$) produces a larger dilepton $v_2$ across the mass range, providing an independent confirmation of the $\eta/s$ driven pattern. Taken together, the three dilepton observables, the invariant mass spectrum, the effective temperature, and the elliptic flow form a mutually consistent picture in which the early stage modifications induced by the $s_{\mathrm{factor}}$-$\eta/s$ retuning leave measurable imprints that hadronic observables cannot access.

\section{Spacetime origin of the electromagnetic sensitivity}\label{sec:spacetime}

The preceding sections showed that electromagnetic observables could distinguish between different minijet contributions in scenarios where various hadronic observables could not. To better understand the changes in the plasma's spacetime profile and its effect on photons, we decompose the thermal photon emission by computing the differential yield $d^3N/(dy\,dT\,d\tau)$ integrated over $p_T \in [0.2, 4.0]$~GeV, as a function of the local medium temperature $T$ and proper time $\tau$. This differential quantity reveals which regions of the fireball evolution contribute most to the photon yield.

\begin{figure}[t]
\centering
\includegraphics[width=0.55\textwidth,height=1.\textheight,keepaspectratio]{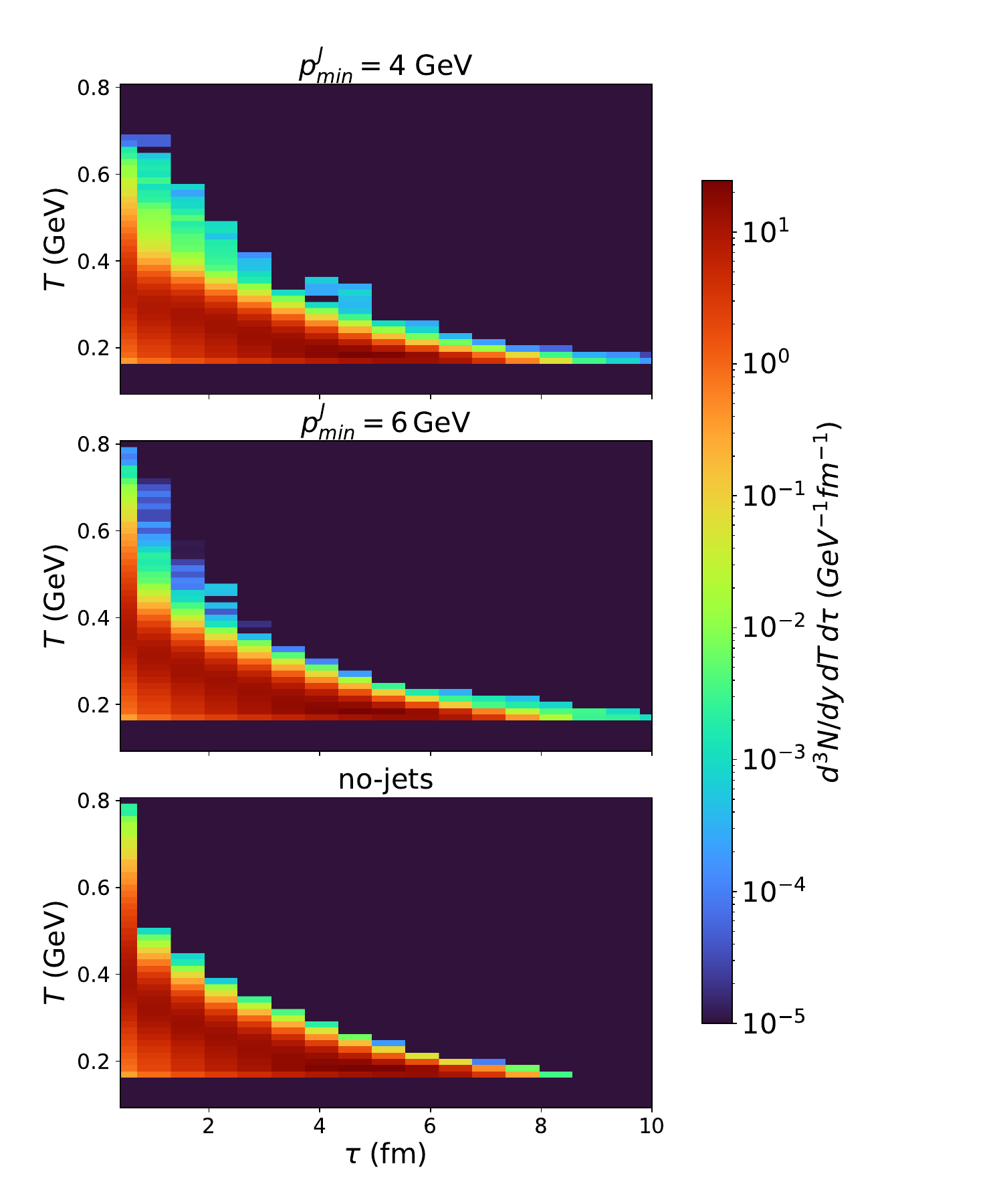}
\caption{Differential thermal photon emission $d^3N/(dy\,dT\,d\tau)$ versus medium temperature $T$ and proper time $\tau$, for three $\pjm$ scenarios and centrality class 10-20\%. The high $T$, early $\tau$ region, where the scenarios differ most strongly.}
\label{fig:heatmap}
\end{figure}

Figure~\ref{fig:heatmap} displays this decomposition for three $\pjm$ scenarios. The emission is concentrated along a ridge that traces the cooling trajectory of the medium: high $T$ at early $\tau$, decreasing at later times. Two features are immediately apparent.

First, comparatively, the no-minijets case exhibits a more pronounced peak in the high-$T$, early-$\tau$ region of the $(T,\tau)$ plane. As the minijet contribution increases, this peak becomes less prominent, while photon production at later times is enhanced relative to the no-minijets case. The thermal emission rates, scaling as high powers of the temperature, are largest in the high-$T$ regime.

Second, it is precisely this early, hot region where the scenarios differ most strongly. The no-jets baseline, with the largest $s_{\mathrm{factor}}$, produces the hottest initial medium and correspondingly the brightest early time emission. As $\pjm$ is lowered, the reduced $s_{\mathrm{factor}}$ yields a cooler initial state. At later times ($\tau\gtrsim 4$-$5$~fm) and lower temperatures ($T\lesssim 200$~MeV), the panels become progressively more similar across scenarios, because the retuned $\eta/s$ has been chosen to match the same hadronic final state and the medium has lost memory of its distinct early stage initial conditions.

This pattern explains the hierarchy of discriminating power among the observables. Hadronic observables, which freeze out at late times and low temperatures, sample the region of the $(T,\tau)$ plane where all scenarios have converged, hence the degeneracy. Direct photons and IMR dileptons, by contrast, are weighted toward the early, hot region where the scenarios are most distinct.

\section{Discussion and outlook}\label{sec:discussion}
We have performed a consistent study of electromagnetic probes within a concurrent minijet-hydrodynamics framework built on IP-Glasma\,+\,PYTHIA\,+\,MUSIC\,+\,UrQMD. By varying the minijet transverse momentum threshold $\pjm$ from $10$~GeV down to $4$~GeV and including a no-jets baseline, we constructed six scenarios for Pb$+$Pb collisions at $\sqrt{s_{NN}}=2.76$~TeV, each calibrated to reproduce the same charged hadron multiplicity and elliptic flow measured by ALICE. Despite a factor of ${\sim}\,2.5$ variation in the specific shear viscosity $\eta/s$ across scenarios, the effect on soft hadronic observables considered here can largely be absorbed here by the retuning of two parameters. 

Electromagnetic observables can lift this degeneracy. The direct photon transverse momentum spectrum shows a systematic decrease in the thermal yield with decreasing $\pjm$, and the direct photon elliptic flow $v_2$ exhibits a clear ordering, with lower $\pjm$ (smaller $\eta/s$) producing larger $v_2$, though current ALICE uncertainties encompass all scenarios. The minijets are produced isotropically, requiring a lower value of $\eta/s$ to describe the hadronic data. This reduction in $\eta/s$, in turn, leads to a larger photon $v_2$. The thermal dilepton invariant mass spectrum in the intermediate mass region ($1\lesssim M\lesssim 3$~GeV) decreases monotonically with decreasing $\pjm$, and the effective temperature $\Teff$ extracted from a fit to this spectrum decreases by ${\sim}\,45$~MeV from the no-jets baseline to $\pjm=4$~GeV. 

A spacetime decomposition of the thermal photon emission into temperature and proper time identifies the origin of this sensitivity. The scenarios differ most strongly in the high $T$, early $\tau$ region of the fireball, which dominates the electromagnetic yield but is invisible to hadronic observables that freeze out at late times and low temperatures.

The present calculation does not include photons produced by the minijets themselves as they traverse the medium. Interactions between the propagating minijet partons and the thermal constituents of the QGP provide an additional source of electromagnetic
radiation~\cite{Turbide:2005fk}. These jet-medium photon channels are a natural next layer of the present framework. Because the number of minijets grows rapidly as $\pjm$ is lowered, their contribution is itself scenario dependent and would add a direct electromagnetic signal of the minijet population.

Heavy quarks and jets offer a complementary hard probe of the same modified background. Charm and bottom quarks are produced in the initial hard scatterings and diffuse through the medium, with their energy loss and flow controlled by the local temperature and velocity fields, precisely the quantities that the concurrent minijet source term reshapes. Evolving heavy quarks through the retuned hydrodynamic backgrounds of Table~\ref{tab:params} would test whether heavy flavor observables, such as the $D$-meson $R_{AA}$ and $v_2$, inherit a sensitivity to the minijet-$\eta/s$ interplay analogous to that found here for thermal photons and dileptons. We can expect a similar effect for hard jets as they traverse a minijet-modified background.

Dilepton polarization, which is sensitive to pressure anisotropy~\cite{Coquet:2023wjk} and is characterized by the angular coefficient $\lambda_\theta$~\cite{Wu:2024vyc,Gao:2026vxs} as a function of invariant mass and $p_T$, has recently been proposed as an independent probe of the early stage dynamics. Since the minijet source term introduces additional momentum space anisotropies and fluctuations into the evolving medium, the six scenarios studied here should also leave an imprint on $\lambda_\theta$, offering a further discriminating observable within the same framework.

Finally, this study only modified two model parameters ($s_{\mathrm{factor}}$ and $\eta/s$). A more comprehensive study can perform a global fit with a minijet-modified hydro, where $\pjm$ is an additional model parameter. This extension can include a larger range of hadronic observables and will be able to distinguish which observables, if any, are responsive to presence of minijets. Over the last decade, Bayesian tools have been extensively used to calibrate the models against the heavy-ion collisions experimental data. The calibrations have often been done on soft hadronic data \cite{Bernhard:2019bmu,Nijs:2020ors,Parkkila:2021tqq,JETSCAPE:2020shq} or on hard hadronic jets or heavy flavor probes \cite{JETSCAPE:2021ehl,Xie:2022ght,Ehlers:2024miy} separately. More recently, efforts have begun to move beyond separate soft- and hard-sector calibrations toward unified Bayesian analyses in which uncertainties and parameters from both sectors are treated consistently. Electromagnetic probes can also be added to such calibrations. A minijet-modified hydro will be a useful addition as these efforts. This will help us leverage electromagnetic and hard probes to calibrate the bulk properties of the medium. As shown in this work, electromagnetic probes can potentially be used to identify the right minijet threshold ($\pjm$) which will impact the extracted transport properties. The same could be true for heavy flavor and jet observables, though that needs to be explicitly demonstrated in future work.

In summary, our study establishes that direct photons and intermediate mass dileptons carry a distinct imprint of the early, hot stage that is erased in the hadronic final state. When combined with a hadronic calibration, electromagnetic probes can help calibrate soft model parameters and give us a clearer picture of how minijets and hydrodynamics evolve concurrently in the quark-gluon plasma.

\begin{acknowledgments}
The computations in this work were performed on the VIKRAM HPC cluster at NISER, Bhubaneswar. S.B. thanks Amaresh Jaiswal and Bedangadas Mohanty for useful discussions and acknowledges financial support from the Department of Atomic Energy (DAE), Government of India. M.S. and J.-F.P. are supported by the U.S. Department of Energy, Office of Science under Award Number DE-SC-0024347 and by Vanderbilt University. D.P. is supported by the Spanish Ram\'on y
Cajal fellowship RYC2023-044989-I. This work was supported in part by the Natural Sciences and Engineering Research Council of Canada (NSERC) [SAPIN-2026-00047 (C.G.) and SAPIN-2024-00026 (S.J.)]. This research used resources of the National Energy Research Scientific Computing Center (NERSC), a U.S. Department of Energy Office of Science User Facility operated under Contract No. DE-AC02-05CH11231. 
\end{acknowledgments}

\bibliography{refs}
\bibliographystyle{apsrev4-2}

\end{document}